\documentstyle[12pt]{article}

\def\refs{\leftskip=.3truein\parindent=-.3truein}
\def\unrefs{\leftskip=0.0truein\parindent=20pt}

\title {The Sign of Four: A new class of cool non-radially pulsating stars?
\thanks{To be published in: {\em Comments on Astrophysics},  1993}}

\author {Kevin Krisciunas}
\date { }

\begin{document}
\maketitle

\vspace{-1 cm}

\begin {center}

  Joint Astronomy Centre \\
  660 N. A'ohoku Place \\
  University Park \\
  Hilo, Hawaii 96720 USA

\end {center}

\begin {abstract}

In this paper we discuss four early F-type variable
stars whose periods are an order of magnitude slower than
known pulsators of comparable luminosity.  They cannot be
stars undergoing simple radial pulsations.  For one or more
of these stars we can discount the possibility that the
variability is due to rotational modulation of star spots,
interactions with (or tidal distortions by) a close
companion, or obscuration by a rotating lumpy ring of dust
orbiting the star.  They are certainly not eclipsing
binaries.  The only possibility left seems to be non-radial
pulsations, though this explanation involves difficulties of
its own.  If they are indeed pulsating stars exhibiting non-
radial gravity modes, they would be the first stars on the
cool side of the Cepheid instability strip in the
Hertzsprung-Russell Diagram to be so identified.

\end{abstract}

\parindent=0pt


\vspace {5 mm}

``[W]hen you have eliminated the impossible, whatever remains,
{\em however improbable}, must be the truth."

\parindent=40pt

-- Sir Arthur Conan Doyle, {\em The Sign of Four}, chapter 6

\parindent=20pt

\begin {center}

{\bf Introduction}

\end {center}

Once upon a time, a colleague of mine set some
undergraduates to observing a particular variable star, using
not one or two, but {\em five} comparison stars.  As luck would
have it, all five of the comparison stars exhibited
variability of differing periods, and the only constant star
during the time of the observations seemed to be the
``variable" star itself!  My rule of thumb is that every star
in the winter sky is variable with the exception of $\phi$\footnotemark[2]
Orionis and BS 1561, and if they too are shown to be
variable, I shall give up astronomical photometry.

I have come to this conclusion after noting that another
region of the HR Diagram has apparently bitten the dust as
far as being a good place from which to choose ``safe"
comparison stars.

\begin {center}

{\bf Data on Four Stars}

\end {center}

My particular nemesis has been an otherwise normal F0 V
star, 9 Aurigae, which has exhibited photometric variability
ranging nearly 0.1 magnitude over the past 6 years.  As
discussed in a recent paper\footnotemark[1], 9 Aur A (the
V $\approx$ 5.0 primary of
a multiple star system) does {\em not} have composition anomalies,
evidence of strong magnetic fields, a close hot companion
(from IUE data), or a close cool companion or circumstellar
ring of dust (from infrared UKIRT and IRAS data).  Its
period(s) of variability allow for the possibility that the
variability is due to rotational modulation of star spots,
but it has been noted\footnotemark[2] that stars earlier than F7 V are too
cool to exhibit extensive spot areas.

Otto Struve would have said that if you study any star
long enough, you would uncover astrophysical anomalies,\footnotemark[3] and
it seemed that 9 Aur was variable without any standard
explanation.  That is of no great importance if the
variations are a few hundredths of a magnitude and it only
pertains to a single star, 9 Aur.  But very slowly I have
learned of other stars with photometric peculiarities similar
to 9 Aur.  In Table I we summarize certain characteristics of
four bright stars, all of which are early F stars on or just
above the main sequence.  Undoubtedly, other stars of
comparable temperature and luminosity will be found to vary
similarly in brightness.

Now F stars are the transition case between hot, early-
type stars which have radiative envelopes, and cool, late-
type stars, which have deep convective envelopes.  The four
stars in Table I are at or just outside the cool edge of the
Cepheid instability strip.  Breger\footnotemark[4] points out, however, that
the ``edges" of the instability strip are not sharp edges -- $\delta$
Scuti stars can be found outside the edges.  Metallic line A-
type stars and others with ``peculiar" spectra (the so-called
Am and Ap stars) can be photometrically variable, but very
few stars of this general spectral type with {\em normal} spectra
found {\em outside} the instability strip are variable (see Fig.
1).

It is well known that the most luminous stars in the
Cepheid instability strip have the longest periods, and that
as one proceeds down the instability strip to less luminous
and hotter stars, the periods monotonically decrease.
Roughly speaking, the characteristic periods of $\delta$ Scuti stars
(which are of spectral type A and F and of luminosity class
IV to V) are 1 to 3 hours.\footnotemark[5]   Assuming for a moment that the
stars in Table I are $\delta$ Scuti stars, and assuming the period-
luminosity relation given by Breger\footnotemark[4], from the absolute
visual magnitudes, the characteristic periods of the stars
should be 42 to 95 {\em minutes}.  The observed periods are roughly
too long by factors of 5, 20 and 40 for HD 96008, $\gamma$ Dor, and 9
Aur, respectively, so it is unlikely that the stars are
undergoing simple radial pulsations.

The relevant papers on HD 96008, $\gamma$ Doradus, HD 164615
and 9 Aur have a certain repetitive ring to them, because the
authors discuss possible reasons for the variability of these
stars and keep eliminating possibilities until almost none
are left.  While the rotational rate of HD 96008 has not been
measured, it is unlikely that spot modulation is the
explanation.  A period of 0.31 days and a stellar radius of 2
R$_\odot$ implies an equatorial rotational speed of about 330
km/sec, which is unlikely for a star that has sharp lines in
its spectrum.\footnotemark[13]  Also, there is the
constraint\footnotemark[2] that the star
is too hot to have extensive star spot areas.  Furthermore,
rotating spotted stars often show evidence for active
chromospheres, and none of these stars, to my knowledge,
shows such evidence.

All four stars exhibit smooth variations of brightness,
so we can rule out the possibility that they are {\em eclipsing}
binaries.  (The primary dip in the light curve of an
eclipsing binary occurs when the brighter star is being
hidden by the fainter star.  The duration of the primary
eclipse is usually short compared to the time between
eclipses.)

Apparently, the period of HD 96008 is stable for years,
and HD 164615 apparently also has a single, stable period. $\gamma$
Doradus, however, has two periods near 3/4 of a day.

9 Aurigae is less well behaved.  In February of 1992 an
observing campaign was carried out by a number of observers
situated at sites differing widely in longitude.  The reason
for this, of course, is to eliminate the 1-day alias in the
power spectrum of the photometry.  This is particularly
important for stars that vary on a time scale comparable to
one day.  Our results\footnotemark[1] were not entirely successful.  The
resultant power spectrum of the data gave a number of peaks
that could be interpreted as the effect of aliases of two
principal periods, 1.277 days and 2.725 days.

Since work on that paper was completed I have received
from Luis Balona a piece of software that allows one to solve
for the amplitudes and phases of periodic signals in a data
set, providing that one knows the frequencies.  If we
reanalyze the 1992 data set and define the periodic signals
by
\begin{center}

$\rm{V}_i = A_i \rm{cos} 2 \pi ( \rm{f}_i \rm{t} + \phi_i )$  ,

\end{center}

\parindent=0pt

we find, for $\rm{f}_1$ = 0.782986, $\rm{A}_1$ = 0.0088 $\pm$ 0.0013 mag,
$\phi_1$ = -0.1650 $\pm$ 0.0266, and for $\rm{f}_2$  = 0.367025,
$\rm{A}_2$  = 0.0094 $\pm$ 0.0014 mag, $\phi_2$  = 0.3780 $\pm$ 0.0237, with
t = HJD - 2448650.0.

\parindent=20pt

Subtracting these two periodic signals from the data set
gives the power spectrum shown in Fig. 2.  Thus, there could
be a third sinusoid with frequency 2.8628 day$^{-1}$ (period =
0.349 days).

If we assume that there are {\em three} bona fide frequencies
present, the resultant amplitudes are 0.0088, 0.0092, and
0.0079 magnitudes for the 1.277, 2.725, and 0.349 day
periods, respectively.  A beat pattern amongst the three
frequencies would give a range of 0.052 magnitude (= 2 times
the sum of all three amplitudes), which is about the mean
range of the available photometry.  (Guiding errors and
variations in atmospheric transparency make the {\em observed}
range larger than the ``true" variations of the star.)

The appearance of the power spectrum in Fig. 2 (and more
recent data discussed below) increases my confidence that the
1.277 day period in particular is a true period.  If I assume
for a moment that the 1-day alias of the 1.277 day period
(namely, 4.608 days) is the true period, and subtract that
and the 2.725 day period from the data set, the resultant
power spectrum is still a forest of lines.

In the winter of 1993 Ed Guinan obtained more
differential photometry of 9 Aur vs. BS 1561, using a 25-cm
robotic telescope at Mt. Hopkins, Arizona.  The data of 31
January through 26 February 1993 UT fold very nicely with the
1.277 day period (see Fig. 3).  An analysis of the 1993 data
similar to that done for the 1992 data yields amplitudes of
0.0218 $\pm$ 0.0009, 0.0022 $\pm$ 0.0007, and 0.0023 $\pm$ 0.0008 mag,
assuming periodic components with periods of 1.277, 2.725 and
0.349 days.  Thus the amplitude {\em ratio} of the two principal
frequencies in February 1992 was 1:1, but by February 1993
changed to 10:1.  Even more confusing is that photometry
obtained by Guinan in early March of 1993, and data obtained
by me in late March of 1993 does not phase up with the light
curve shown Fig. 3, being out of phase by 0.6 cycles.  At
face value 9 Aur appears able to vary at a given period, then
stop altogether for a sizable fraction of a cycle, then
resume varying with the same period.  If this is a real
effect, it would make it impossible to tie together data sets
from different years.  Suffice it to say that the photometric
behavior of 9 Aur is the least understood of the four stars
in Table I.

\begin {center}

{\bf Non-radial Pulsations}

\end {center}

Pulsation theory confidently describes the behavior of
stars in the Ceph\-eid instability strip undergoing simple
radial pulsations.  The least dense, largest stars
(supergiant Cepheids) pulsate most slowly, and the densest,
smallest stars (DA white dwarfs) pulsate most rapidly.
Pulsations with shorter periods than the fundamental radial
mode are attributed either to overtones of the fundamental
radial mode, or to non-radial p-modes, where ``p" stands for
``pressure".  In order to have variations much {\em slower} than the
fundamental radial mode, one must appeal to non-radial g-
modes, where ``g" stands for ``gravity".

Non-radial pulsations are described mathematically by
spherical harmonics, such as those used in classic
electrodynamics and quantum mechanics texts.  A spherical
harmonic for a star is described by integral indices {\em k}, {\em l},
and {\em m}, where {\em k} is the radial overtone number,
{\em l} is the
``degree" of the harmonic, and {\em m} is the azimuthal index.  The
{\em l} index gives the total number of nodal planes slicing the
star's rotation axis.  The {\em m} index indicates the number of
these nodal planes that contain the star's rotational axis.
For every {\em l} there are 2{\em l}  + 1 possible values of
{\em m} (ranging
from -{\em l} to +{\em l}).  By convention positive {\em m} values signify
waves going in the direction of the star's rotation, while
negative {\em m} values signify waves running opposite that of the
direction of rotation.

For example, an {\em l} = 1, {\em m} = 0 harmonic is like a fat man
pulling his belt over his belly again and again.  The belt
does not stay up for long and his belly keeps coming over the
top.  In the case of a star the northern hemisphere slightly
swells and gets brighter while the southern hemisphere
simultaneously shrinks and gets slightly fainter.  Then the
process reverses.  If we were observing such a stellar mode
from the {\em pole on} direction, the star would vary cyclically in
brightness, and its radial velocity would also vary.

For the case of {\em l} = 1, {\em m} = $\pm$1, there are two waves
running around the longitude lines of the star, with one wave
running the direction of the star's rotation, and,
simultaneously, another wave running in the opposite
direction.  In this case if the star were viewed from the
{\em side}, we would observe brightness variations and radial
velocity variations, while there would be no observed
variations due to the {\em l} = 1, {\em m} = 0 mode.

The cases of {\em l} = 1, 2, 3, and 4 are shown graphically
by Winget and Van Horn.\footnotemark[14]  The case of {\em l} = 6 is shown by
Nather and Winget.\footnotemark[15]
Pesnell\footnotemark[16] describes a movie that
illustrates various non-radial pulsations.
Cox$^{5,17}$ gives
more general descriptions.  The effect of the inclination of
the star's rotation axis to the observer's line of sight on
the observed lines in the power spectrum of a non-radially
pulsating star is described by Pesnell.\footnotemark[18]

Simply put, a non-radial pulsator can exhibit many modes
simultaneously, and whether or not one observes brightness
and radial velocity variations depends on the mode ({\em l} and
{\em m}), the angle of the star's rotation axis to the line of
sight, and its limb darkening.  Finally, rotation will split
the lines for a given multiplet.  For a given degree {\em l}, the
largest positive value of the azimuthal index ({\em m} = {\em l}) has
the highest frequency (shortest period).  However, it must be
noted that a star can damp out certain modes, so that one
does not necessarily observe what theory predicts.

The ``classic" {\em slow} non-radial pulsator (i.e. a star
undergoing non-radial g-modes) is 53 Persei, classified B4
IV,\footnotemark[19] which has a surface temperature of about
16,000 K.\footnotemark[20]

Other stars that are presumed to exhibit non-radial g-
modes are all very hot.  In the case of the DOV star PG 1159-035,
which has a surface temperature of 123,000 K, the power
spectrum shows over 100 frequencies, usually appearing as
triplets and quintuplets.  Winget {\em et al.}\footnotemark[21]
conclude that the
star is exhibiting non-radial g-modes with {\em l} = 1 and {\em l} = 2,
which are stimulated by radial overtones of order 20.  (The
spacings between multiplets tell that more than one radial
overtone is in action, but one cannot specify exactly what
values of {\em k} are involved.)

We note that PG 1159-035 and 53 Per are not found in the
{\em Cepheid} instability strip, and that DB white dwarfs (with
temperatures of 30,000 K) also show non-radial g-modes.
There are no A, F, or G main sequence or subgiant stars that
have been shown to exhibit non-radial g-modes.  (The Sun,
exhibits non-radial p-modes, and, I am told, is {\em suspected} of
having non-radial g-modes.)

Other than the anomalously long periods, what evidence
do we have that the stars in Table I might be non-radial g-mode
pulsators?  First of all, $\gamma$ Doradus exhibits two closely
spaced periods.  Since the whole surface of a star usually
has one ``characteristic" rotational period, the two periods
of $\gamma$ Doradus are evidence against rotational modulation of
some kind of spots on that star's surface.  (The Sun, of
course, has a rotational period that is a function of
latitude.  In the case of $\gamma$ Doradus it would take spots
confined to specific latitude bands, with differing
rotational periods, to produce the observed effect.  Along
these lines, Smith\footnotemark[22] discusses the role of differential
rotation for non-radial pulsations of Be stars.)

Abt, Bollinger, and Burke\footnotemark[11] measured 13 radial
velocities of HD 164615, which ranged 7.4 km/sec, but given
the star's line widths (owing to its reasonably rapid
rotation rate) and the internal errors, they concluded that
the star's radial velocity was constant.

In ref. 8 we show a graph of 28 radial velocities of 9
Aur (with typical errors of $\pm$0.2 km/sec) obtained by Helmut
Abt over a two hour period, indicating essentially a constant
radial velocity.  There was clearly no evidence for
pulsations like those of a normal $\delta$ Scuti star, with a period
of 1-3 hours.  However, as mentioned in ref. 1, radial
velocities of 9 Aur in the literature, if taken at face
value, indicate a range of 6 km/sec.  As a result, Roger
Griffin kindly obtained 20 radial velocities in February in
March of 1993, using a CORAVEL instrument at Haute Provence.
While the fully reduced data will not be available from the
CORAVEL data base for some time, it can be said that the
radial velocity of 9 Aur does indeed seem to range 6 km/sec,
and vary on a daily basis.  What we need next are radial
velocities obtained at a rate of one per hour for several (or
many) nights running.

Another type of evidence in favor of non-radial
pulsations comes from high resolution spectroscopy.  Non-
radial pulsators will exhibit changing line profiles, and the
type of changes is indicative of the mode of non-radial
pulsations.\footnotemark[23]

Coupry and Burkhart\footnotemark[24] recently showed that 9 Aur can
exhibit asymmetric line profiles with blue wings, but they
obtained only two spectra.  While obtaining his radial
velocities of 9 Aur in February and March of 1993, Griffin
noted that the lines were sometimes asymmetric.  When the
fully reduced data are available from the CORAVEL data base,
we will be able to check Griffin's real-time impression more
quantitatively.

Balona {\em et al.}\footnotemark[25] relate that some spectra
of $\gamma$ Doradus by John
Hearnshaw exhibit varying line profiles indicative of an {\em l} =
2 non-radial pulsator.

I apologize for all the references to sporadically taken
data, unreduced data, unpublished data, and comments from
private communications that may not be proven true.  But
given the anomalously slow periods of the stars in Table I,
indications of line profile variations in two of them, and
evidence for radial velocity variations in one of them, the
presently available data are consistent with the notion that
one or more of our stars in Table I are non-radial g-mode
pulsators.  These would be the first cool stars ($\approx$ 7000 K) so
identified.

\begin {center}
{\bf Discussion}
\end {center}

One of the biggest problems with understanding slow non-
radial pulsators, be they hot stars like 53 Per or cooler
stars like the F stars discussed here, is the {\em paucity} of
frequencies in the light curves.  How can a star be excited
to pulsate from its {\em k}-th overtone, but not show frequencies
from the ({\em k}-1)-th and ({\em k}+1)-th?  Could it be that density
gradients or composition gradients within the star, coupled
with its core rotation rate, damp out all but one or two of
the pulsational modes?\footnotemark[22]

To help substantiate the case that the stars in Table I
are non-radial pulsators, on the observational side we need
the following:

1) Coordinated photometry by multiple observers situated
around the globe.  Unfortunately, the stars in question are
much brighter and have periods considerably slower than typical
targets of the Whole Earth Telescope.\footnotemark[15]

2) Concentrated batches of radial velocities.  This will
be more easily accomplished for 9 Aur and HD 96008, which
have sharp lined spectra.  (Also, it would be a good idea to
measure the rotation rate of HD 96008.)

3) Sequences of high resolution line profiles.  This
will be more easily accomplished for $\gamma$ Doradus and HD 164615,
which have reasonably broad lines.

4) It would be good if we knew of more than the four
stars listed in Table I.  How many measurements of early-F
dwarfs are there that photometrists have thrown out of the
data reduction because occasional observations were a few
hundredths of a magnitude ``wrong"?  Photometrists should scan
their memory banks and their observing notebooks for evidence
of suspicious candidates, following up with more photometry
to substantiate whether candidates are bona fide variables.
Some will be $\delta$ Scuti stars with periods of 1-3 hours.  Some
might have periods like our four stars described here.

On the theoretical side, we need realistic model
atmospheres of F dwarf stars, and a better understanding of
non-radial pulsation theory.  It is understood that both of
these subjects are non-trivial.

I find it simultaneously inspiring, humbling, and
frustrating that stars only a little more massive than the
Sun are variable in brightness but exhibit no other obvious
peculiarities.  We say that we understand main sequence
stars, but here is yet another phenomenon that is
unexplained.

Invoking the idea of Sherlock Holmes quoted at the start
of this article, for the time being we might assume that the
stars discussed here are non-radial pulsators.  But if we can
prove that these stars are {\em not} non-radial pulsators, then we
will have run out of sensible explanations.  In that case we
would have to admit that our four stars have conspired
successfully to confound us completely, and we must put off
solving the case until our understanding of physics is much
greater.

\begin {center}
{\bf Acknowledgments}
\end{center}

I gratefully acknowledge recent discussions I have had
with Jay\-mie Mat\-thews, Roger Griffin, Patricia Lampens, Luis
Balona, Ed Guinan, Myron Smith, and Dean Pesnell.  I hope I
have not misrepresented any of their comments.
I also benefited from the use of the SIMBAD data
retrieval system, data base of the Strasbourg, France,
Astronomical Data Center.

\newpage
\begin {center}
{\bf References}
\end {center}

\refs

1. K. Krisciunas, C. Aspin, T. R. Geballe, H. Akazawa, C. F.
   Claver, E. F. Guinan, H. J. Landis, K. D. Luedeke,
   N. Ohkura, O. Ohshima, and D. R. Skillman, Monthly
   Notices R. A. S., in press (1993)

2. M. S. Giampapa and R. Rosner, Astrophys. J. Lett. 286,
   L19 (1984)

3. K. Krisciunas, Biographical Memoirs (National Academy
   Press, Washington, D. C.), 61, 351 (1992)

4. M. Breger, PASP 91, 5 (1979)

5. J. P. Cox, {\em Theory of Stellar Pulsation} (Princeton Univ.
   Press, Princeton, NJ, 1980)

6. A. W. J. Cousins, Observatory 112, 53 (1992)

7. K. Krisciunas and E. Guinan, IBVS 3511 (1990)

8. K. Krisciunas, E. F. Guinan, D. R. Skillman, and
    H. A. Abt, IBVS 3672 (1991)

9. P. Lampens, Astron. Astrophys. 172, 173 (1987)

10. J. M. Matthews, Astron. Astrophys. 229, 452 (1990)

11. H. A. Abt, G. Bollinger, and E. W. Burke, Jr., Astrophys.
   J. 272, 196 (1983)

12. S. M. Rucinski, PASP 97, 657 (1985)

13. According to Balona, as quoted in Lampens (ref. 9) above

14. D. E. Winget and H. M. Van Horn, Sky Tel. 64, 216 (1982)

15. R. E. Nather and D. E. Winget, Sky Tel. 83, 374 (1992)

16. W. D. Pesnell, Amer. J. Physics 53, 579 (1984)

17. J. P. Cox, PASP 96, 577 (1984)

18. W. D. Pesnell, Astrophys. J. 292, 238 (1985)

19. R. J. Buta and M. A. Smith, Astrophys. J. 232, 213 (1979)

20. M. A. Smith and R. J. Buta, Astrophys. J. Lett. 232,
    L193 (1979)

21. D. E. Winget and 31 coauthors, Astrophys. J. 378, 326
    (1991)

22. M. A. Smith, in {\em Pulsation and Mass Loss in Stars}, ed.
    R. Stalio and R. A. Willson (Kluwer, Dordrecht, 1988),
    pp. 251-274

23. C. Aerts and C. Waelkens, Astron. Astrophys. 273, 135
    (1993)

24. M. F. Coupry and C. Burkhart, Astron. Astrophys. Suppl.
    95, 41 (1992)

25. L. A. Balona, J. B. Hearnshaw, C. Koen, A. Collier, I. Machi,
    M. Mkhosi, and C. Steenberg, Monthly Notices R. A. S., in
    press (199[3])

26. W. Gliese, {\em Catalogue of Nearby Stars}, Veroeffentl.
    Astron. Rechen-Inst. Heidelberg, No. 22 (1969)

\unrefs

\newpage

\begin {center}
{\bf Figure Captions}
\end {center}

1. Incidence of variability of stars with known normal
spectra.  From Breger (ref. 4).  The positions of three of
the four stars in Table I are indicated by X's.  (HD 164615
does not have a color or absolute magnitude available.  We
adopt Gliese's\footnotemark[26] absolute magnitude for 9 Aurigae.)

\vspace{5 mm}

2. Power spectrum of 9 Aurigae data obtained from 31 January
to 10 February 1992 UT, after subtracting 1.227-day and
2.725-day signals.  This indicates a possible third signal
with frequency 2.8628 day$^{-1}$ (period 0.349 days).

\vspace{5 mm}

3. Differential photometry of 9 Aurigae vs. BS 1561 by Guinan
in early 1993, using a 25-cm robotic telescope at Mt.
Hopkins, Arizona.  The data are folded by the 1.277 day
period.

\end{document}